\documentclass[prl,twocolumn,superscriptaddress,nofootinbib]{revtex4}

\usepackage{amsfonts}
\usepackage{amsmath}
\usepackage{amssymb}
\usepackage{amsthm}
\usepackage{bm}
\usepackage{dcolumn}
\usepackage{epsfig}
\usepackage{graphicx}
\usepackage{graphics}

\newcommand{\GW}{{\mbox{\tiny GW}}}

\newcommand\be{\begin{equation}}
\newcommand\ba{\begin{eqnarray}}
\newcommand\ee{\end{equation}}
\newcommand\ea{\end{eqnarray}}

\begin{document}

\title{Modeling Extreme Mass Ratio Inspirals within the Effective-One-Body Approach}

\author{Nicol\'as Yunes}
\affiliation{Department of Physics, Princeton University, Princeton, NJ 08544, USA.}

\author{Alessandra Buonanno}
\affiliation{Maryland Center for Fundamental Physics, Department of Physics, University of Maryland, College Park, MD 20742, USA.}

\author{Scott A.~Hughes}
\affiliation{Department of Physics and MIT Kavli Institute, 77 Massachusetts Avenue, Cambridge, MA 02139, USA.}

\author{M.~Coleman Miller}
\affiliation{Maryland Astronomy Center for Theory and Computation, Department of Astronomy, University of Maryland, College Park, MD 20742, USA.}

\author{Yi Pan}
\affiliation{Maryland Center for Fundamental Physics, Department of Physics, University of Maryland, College Park, MD 20742, USA.}

\begin{abstract}
We present the first models of extreme-mass-ratio inspirals within the
effective-one-body (EOB) formalism, focusing on quasi-circular orbits
into non-rotating black holes.  We show that the phase difference and
(Newtonian normalized) amplitude difference between analytical EOB and
numerical Teukolsky-based gravitational waveforms can be reduced to $\lesssim
10^{-1}$ rads and $\lesssim 2 \times 10^{-3}$, respectively, after a
2-year evolution. The inclusion of post-Newtonian self-force terms in
the EOB approach leads to a phase disagreement of $\sim 6 \mbox{--}
27$ rads after a 2-year evolution.  Such inclusion could also allow
for the EOB modeling of waveforms from intermediate-mass ratio,
quasi-circular inspirals.
\end{abstract}

\maketitle

{\it{Introduction}}.~Extreme mass ratio inspirals (EMRIs) are the gravitational
wave (GW) driven coalescences of stellar mass compact objects
with supermassive black holes (SMBHs).  When the large black hole's (BH's)
mass is in the range $10^5\,M_\odot - 10^7\,M_\odot$, EMRI waves are
emitted at frequencies well suited to measurement by the planned {\it
Laser Interferometer Space Antenna} (LISA).  Because EMRI events are
expected to be abundant {\cite{2004CQGra..21S1595G}} and will carry
detailed information about strong-field spacetimes near SMBHs
{\cite{Hughes:2006pm}}, they are high-priority targets for LISA
observation.  The intrinsic feebleness of these waves will require
accurate waveform templates to detect and faithfully measure the
signals produced by Nature.  
Because EMRIs can spend thousands of cycles in the close vicinity of the SMBH's
innermost stable circular orbit (ISCO), traditional post-Newtonian expansions
are not well suited to modeling their signals; the binary's orbital speed
is $v/c \sim 0.1 \mbox{--} 0.5$, a regime where traditional post-Newtonian techniques
perform poorly. Numerical models built using BH perturbation theory should be able to
reliably model EMRI signals.  However, the computational cost of
covering the full span of EMRI parameter space (including effects due
to BH spin, non-equatorial orbits, and eccentricity) is likely to be
very high {\cite{2004CQGra..21S1595G}}.

This has motivated us to examine techniques for reliably approximating
these waves at a much smaller computational cost.  The
effective-one-body (EOB) formalism was introduced as a way to
analytically describe the inspiral, merger, and ringdown waves emitted
by comparable-mass BH binaries~\cite{Buonanno99,Buonanno00}.  This
formalism was then extended to higher post-Newtonian (PN)
orders~\cite{Damour00}, spinning BHs~\cite{Damour01,Buonanno06,Damour:2007nc}, small
mass-ratio mergers~\cite{Nagar:2006xv,Damour2007}, and was further improved by
introducing factorized waveforms~\cite{Damour2007,Damour:2008gu}.  By
calibrating a few adjustable parameters in the EOB-dynamics and
waveforms, \cite{Damour:2009kr,Buonanno:2009qa} showed that the phase
and amplitude of the EOB and numerical-relativity waveforms can be
made to agree within the numerical error of the simulations, thus
providing GW detectors with faithful templates.  
In this analysis, we consider calibrating EOB with BH perturbation theory templates in order 
to similarly model EMRI waves.  Such an analysis must be done separately 
from the previous EOB-numerical relativity calibration, because sufficiently long numerical relativity
simulations are currently not available for EMRIs. This is because the large mass ratio in EMRI systems 
leads to tens to hundreds of thousands of detectable cycles within a one-year window using LISA,
thus requiring extremely long simulations that are currently computationally prohibitive.

As a first step, we restrict our models to a small compact object
spiraling along a quasi-circular orbit into a non-spinning
SMBH~\cite{2007CQGra..24..113A}.  Although the assumptions of
circularity and zero spin can and will be relaxed in the future, there
exist astrophysical motivations for this initial choice of binary
configuration.  For example, the tidal separation scenario for EMRIs
\cite{2005ApJ...631L.117M} implies nearly circular but arbitrarily
inclined orbits in the $>10^{-4}$~Hz frequency band relevant for LISA,
and the accretion disk capture picture \cite{1991MNRAS.250..505S,Levin:2003ej,Levin:2006uc}
implies orbits that are both nearly circular and in the equatorial plane of
the SMBH.  In addition, the characteristics of the SMBHs themselves
are uncertain in the $<10^7~M_\odot$ mass range most relevant for
LISA.  In some astrophysical scenarios, the growth of these BHs is
dominated by the accretion of stars moving on random trajectories,
instead of by the accretion of gas disks, thought to be more important
for higher-mass SMBHs \cite{2004ApJ...600..149W}.  Such growth would
lead to ${\hat a}\equiv |{\vec J}|/M^2\ll 1$ (in natural units with
$G=c=1$, which we use throughout this paper), hence the Schwarzschild
(nonrotating) spacetime is a reasonable first approximation.

We now systematically compare EMRI waveforms computed in the EOB
approach to those calculated using BH perturbation theory via the numerical 
solution of the Teukolsky equation~\cite{Teukolsky:1972le,Teukolsky:1973ap,Hughes:1999bq,Hughes:2001jr}.  
As we describe in the remainder of this {\it Letter}, we find that
appropriately calibrated EOB waveforms in fact do an excellent job
modeling waves computed using BH perturbation theory.  This suggests
that the EOB scheme is very likely to be an outstanding tool for modeling EMRI
waves in future LISA data analysis. 

{\it{Analytical and Numerical Modeling}}. 
For a BH binary with masses $m_1$ and $m_2$, we set $M=m_1+m_2$ and $\mu = m_1\,m_2/M = \nu \,M$.
In absence of spins, the motion is constrained to a plane.  Let us introduce Schwarzschild-like coordinates $(r,\Phi)$ (where $r$ is $M$-normalized) centered on the binary's center of mass, as well as their reduced ($\mu$-normalized) conjugate momenta $(p_r, p_\Phi)$.  The non-spinning EOB Hamiltonian then reads~\cite{Buonanno99} 
$H^{\rm real} = M\,\sqrt{1 + 2\nu\,[ (H^{\rm eff} - \mu)/{\mu}]} -M$,
where the effective Hamiltonian is~\cite{Buonanno99,Damour00,Damour2007}
\begin{equation}
  \label{eq:genexp}
  H^{\rm eff} = \mu\,\sqrt{p^2_{r_*}+A (r) \left[ 1 + \frac{p_\Phi^2}{r^2} 
      + 2(4-3\nu)\,\nu\,\frac{p_{r_*}^4}{r^2} \right]} \,.
\end{equation}
We use here the reduced conjugate momentum $p_{r_*}$ to the EOB {\it tortoise} radial 
coordinate $r_*$ because it improves the numerical stability of the code~\cite{Damour2007}. 
The tortoise coordinate is defined via $dr_{*}/dr = \sqrt{D(r)}/A(r)$, where $A(r)$ 
and $D(r)$ are obtained by applying the Pad\'e resummation~\cite{Damour00} to the Taylor-expanded 
forms~\cite{Buonanno99,Damour00}
\begin{eqnarray}
A_{\rm T}(r) &=& 1 - \frac{2}{r}+\frac{2\nu}{r^{3}}+ \left (
\frac{94}{3}-\frac{41}{32}\pi^2\right )\,\nu\,\frac{1}{r^{4}}\,,\\
D_{\rm T}(r) &=& 1 -\frac{6\nu}{r^{2}}+2\nu\, (3\nu-26)\,\frac{1}{r^{3}}\,. 
\end{eqnarray}
The EOB Hamilton equations are written in terms of the reduced 
(dimensionless) quantities $\widehat{H}^{\rm real} \equiv H^{\rm real}/\mu$, 
$\widehat{t} = t/M$~\cite{Buonanno00}: 
\begin{eqnarray}
  \frac{dr}{d \widehat{t}} &=& 
  \frac{A(r)}{\sqrt{D(r)}}\frac{\partial \widehat{H}^{\rm real}}
  {\partial p_{r_*}}\,, \quad \quad \frac{d \Phi}{d \widehat{t}} =
  \frac{\partial \widehat{H}^{\rm real}}  {\partial p_\Phi}\,,
  \label{eq:eobhamonetwo} \\
  \frac{d p_{r_*}}{d \widehat{t}} &=&- 
  \frac{A(r)}{\sqrt{D(r)}}\,\frac{\partial \widehat{H}^{\rm real}}
  {\partial r}\,, \quad \quad \frac{d p_\Phi}{d \widehat{t}}=
  \widehat{\cal F}_\Phi\,,
  \label{eq:eobhamthreefour}
\end{eqnarray}
where $\widehat{\cal F}_\Phi$ is a Pad\'e-resummed radiation-reaction force~\cite{Damour:1997ub,Buonanno00}, 
related to the GW energy dissipation to be defined later.
Initial data is obtained through a mock evolution, 
which is initialized at an initial orbital separation of $100 M$ using 
initial conditions for a quasi-circular inspiral~\cite{Buonanno00}. 

With the EOB inspiral dynamics in hand, we compute the multipole-decomposed 
GW $h_{\ell m}$ ($\ell$ and $m$ refer to spherical harmonics), following the factorized PN 
prescription of~\cite{Damour:2008gu}, which depends directly on orbital quantities. The EOB GW phase is computed by solving $\dot{\Phi}_{\ell m} = - (1/m)\,{\rm Im}[\dot{h}_{\ell m}/{h}_{\ell m}]$. 
Errors in the EOB waveforms arise due to inaccuracies in the numerical
solution of Eqs.~(\ref{eq:eobhamonetwo}) and (\ref{eq:eobhamthreefour})
and inaccurate initial data. We have investigated such sources of error and estimate them 
to be no worse than $\delta \Phi_{22} \lesssim 0.03$ rads in the waveform's phase and 
$\delta h_{22}/h_{22} \lesssim 10^{-7}$ in the normalized amplitude after a 2-year evolution. 
This cumulative error is primarily dominated by the accuracy of the routine used in {\tt Mathematica} to 
solve Eqs.~(\ref{eq:eobhamonetwo}) and (\ref{eq:eobhamthreefour}).

We compare EOB waves with waveforms computed in BH perturbation theory 
by solving the Teukolsky equation.  We use the code described in~\cite{Hughes:1999bq} (modified with
the spectral techniques of~\cite{Fujita:2005}) to construct the Newman-Penrose
curvature scalar $\psi_4$. Our code works in the frequency domain, decomposing $\psi_4$ into  
$\psi_4 = R^{-1}\sum_{\ell m}Z_{\ell m}\,{}_{-2}Y^{\ell m}(\theta,\phi)\,e^{-im\Omega t}$ 
where ${}_{-2}Y^{\ell m}(\theta,\phi)$ is a spin-weight $-2$ spherical harmonic,  
$\Omega$ is the frequency of circular Schwarzschild orbits and $R$ is the distance from 
the center of mass to the observer.  
The amplitude $Z_{\ell m}$ is found by first constructing a Green's function to the
radial Teukolsky equation, and then integrating that Green's function over a source made from the 
stress-energy tensor of the small body orbiting the BH; see~\cite{Hughes:1999bq} for specifics.

The radial Teukolsky equation possesses two asymptotic solutions that determine the behavior of 
$\psi_{4}$ at spatial infinity and near the event horizon. 
In the distant radiation zone, $\psi_4$ is related to the GWs carried away from the system via 
$\psi_4 \to {1}/{2}(\ddot h_+ - i\ddot h_\times)$. Therefore, the solution to the radial Teukolsky equation that describes
purely outgoing radiation at spatial infinity can be used to construct the flux of radiation and the waveform that distant 
observers measure. On the other hand, near the event horizon, $\psi_4$ describes tidal interactions of the BH with the orbiting 
body~\cite{Teukolsky:1973ap}. Thus, the solution to the radial Teukolsky equation that describes purely ingoing radiation at the
horizon can be used to construct the radiation flux  absorbed by the BH. With these fluxes, we can then calculate 
the rate at which the orbital radius changes, $\dot r$, by noting that for slow backreaction the system evolves through 
a sequence of geodesic orbits.

We construct the $\psi_4$ solution on a discrete grid of orbits from $r = 10,000M$ to the Schwarzschild ISCO at $r = 6M$ (in Boyer-Lindquist coordinates), evenly spacing our orbits in $v \equiv \sqrt{M/r}$.  (Since stable circular orbits do not exist for $r < 6M$,
we cannot infer $\dot r$ from $dE/dt$ in this regime.)  Errors in the Teukolsky-based waveforms are dominated by 
truncation of the $(\ell,m)$ sums in $\psi_{4}$ and due to the discretization of the orbital phase space
when the fluxes are cubic-spline interpolated from a discrete adiabatic sequence of geodesic orbits. 
The sums and the discretization are chosen such that the fractional error in the flux is smaller than 
$10^{-10}$~\cite{Hughes:1999bq,Hughes:2001jr}. In practice, in the low velocity region $v < 0.1$, 
we find that the flux is accurate to at least $10^{-13}$. Such an error translates to inaccuracies 
in the GW phase of less than $10^{-2}$ rads over a 2-year evolution.

{\it{Systems, Regions and Models}}. 
To demonstrate the flexibility of the EOB model in matching the Teukolsky-based waveforms, 
we examine two fiducial EMRI systems, labeled system-I and system-II, that sample different regions of the LISA
noise curve.  In both cases, we consider a 2-year long quasi-circular inspiral 
of non-spinning BHs. System-I has $(m_1,m_{2}) = (10^5,10) M_{\odot}$; system-II has $(m_1,m_{2})=(10^6,10) M_{\odot}$. We do not consider lower or higher total mass binaries as they 
would either reach the ISCO outside the LISA optimal sensitivity band (LISA's noise rises sharply above $\sim 10^{-2}$ Hz) 
or lie significantly inside the white-dwarf confusion limit (much below $\sim 0.002$ Hz~\cite{Farmer:2003pa}). System-II ($m_2/m_1=10^{-5}$)
begins at an initial separation $r_{\rm in} \simeq 10.6 M$ 
and terminates at the ISCO, sweeping GW frequencies in the range 
$f_{\GW} \in [1.8 \times 10^{-3},4.4 \times 10^{-3}] \; {\rm{Hz}}$. 
System-I ($m_2/m_1=10^{-4}$) starts at 
$r_{\rm in} \simeq 29.34M$ and terminates at $r_{\rm fin} \simeq 16.1 M$, sweeping frequencies
in the range $f_{\GW} \in [4 \times 10^{-3},10^{-2}] \; {\rm{Hz}}$.
The mass ratios we consider, $(10^{-4}, 10^{-5})$ are two orders of magnitude smaller than those studied in the complementary analyses of~\cite{Nagar:2006xv,Damour2007}.  As such, our in-band signal is dominated by a long inspiral; the contributions of the final plunge, merger, and ringdown, which dominate the signal of~\cite{Nagar:2006xv,Damour2007}, are much less important here.
These choices allow the study of the early and late EMRI dynamics, while guaranteeing the 
GW signal is in the sensitive part of the LISA band. 

We define two EOB models differing in the resummation of the radiation-reaction force in 
Eq.~(\ref{eq:eobhamthreefour}). Using the balance law, we write $\widehat{\cal F}_\Phi=  -F /(\nu \Omega)$, where 
$F$ is the GW energy flux. We use (i) the Pad\'e-approximant to the energy 
flux~\cite{Damour:1997ub,Buonanno:2009qa} $F^{\rm P}= {F}_q^p(v_{\rm pole})$, where $v_{\rm pole}$ 
is an adjustable parameter locating the EOB light-ring, and 
$p+q$ is twice the approximant's PN order [i.e., $(v/c)^{(p+q)}$], and 
(ii) the $\rho$-approximant to the energy flux~\cite{Damour:2008gu}  
${F}^{\rho}= {2}/(16\pi)\sum_{\ell =2}^{\ell=8}\sum_{m=1}^{m=\ell} (m\Omega)^2|Rh_{\ell m}|^2$.
Except when investigating the effect of the self-force, the orbital dynamics are  
computed setting $\nu=0$ in $F$, as well as in $A(r)$ and $D(r)$, i.e., for a Schwarzschild 
BH.

\begin{figure}
 \epsfig{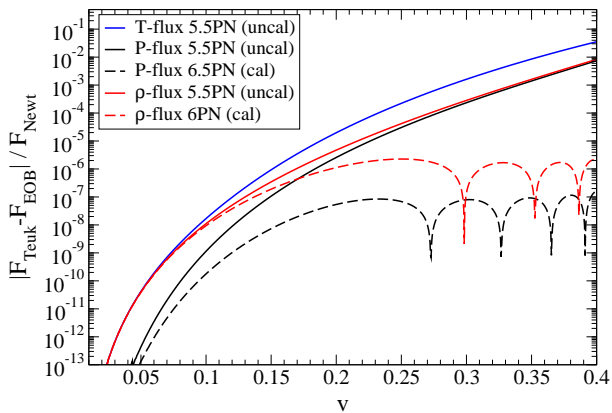} 
 \vspace{-0.4truecm}
 \caption{\label{fig:flux} Absolute value of the difference in the Newtonian normalized Teukolsky and EOB  
 fluxes as a function of orbital velocity.  Calibrating the Pad\'e or
$\rho-$flux improves the agreement by orders of magnitude}
\vspace{-0.5truecm}
\end{figure}

{\it{Results}}. 
Figure~\ref{fig:flux} shows the absolute value of the difference between 
the Newtonian-normalized ($F_{\rm Newt} = 32 \nu^2 v^{10}/5$) Teukolsky and EOB 
({\it uncalibrated} and {\it calibrated}) fluxes as a function of the orbital velocity $v$. 
The Teukolsky flux includes energy both radiated to infinity and absorbed by the BH's event horizon. 
The uncalibrated Pad\'e-flux (${F}_4^7$) and $\rho$-flux are computed through 5.5PN order, 
but in the Pad\'e flux we also add horizon absorption corrections~\cite{Mino:1997bx} and 
set $v_{\rm pole}$ to the Schwarzschild light-ring value. The uncalibrated 
Taylor-flux (i.e., the PN Taylor-expanded flux~\cite{Boyle:2008ge}) 
gives a residual about five times worse than the uncalibrated Pad\'e and $\rho$ fluxes.
The calibrated Pad\'e-flux (${F}_6^7$) is computed through 6.5PN order, including the horizon absorption
corrections, and calibrating $v_{\rm pole}$ and the 6PN and 6.5PN coefficients 
${\cal{F}}_{12}$ and ${\cal{F}}_{13}$; see~\cite{Boyle:2008ge} for details. 
The calibrated $\rho$-flux is computed through 6PN order, without horizon absorption corrections,
and calibrating the 6PN coefficients $c_6^{\rho_{22}}$ in $\rho_{22}$ and the 5PN coefficients 
$c_5^{\rho_{33}}$ in $\rho_{33}$; see~\cite{Damour:2008gu} for details. 
The calibration is here performed via a least-squares fit to the numerical Teukolsky flux. 
For velocities $v \in [0.01,0.1]$ the agreement is better than $10^{-8}$ with a best agreement of 
$10^{-13}$ near $v = 0.01$ for all models.

\begin{figure*}
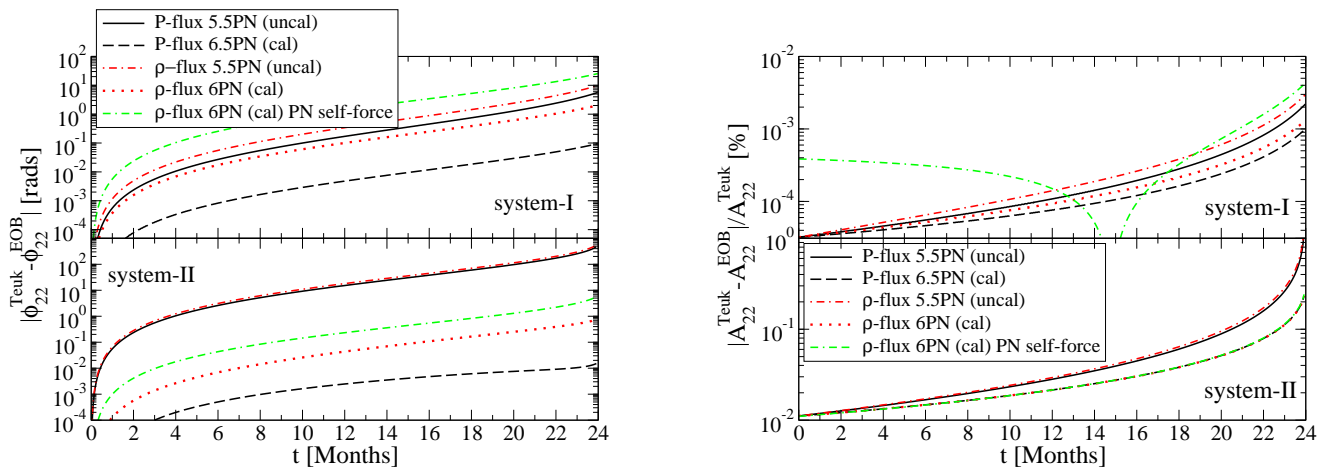

\begin{center}
\begin{tabular}{cc}
 \includegraphics[width=8cm,clip=true]{phase-evolution-sys-combined.eps} \qquad \qquad
  \includegraphics[width=8cm,clip=true]{amp-evolution-sys-combined.eps}  
\end{tabular}
\end{center}
 \vspace{-0.5truecm}
\caption{\label{fig:plot2} 
Absolute value of the dephasing (left) and fractional amplitude difference (right) of the dominant 
GW $(2,2)$ mode as a function of time in months.  Again, with the introduction
of calibrated higher-order terms, the differences are small even over a full
two year coherent integration.}
\vspace{-0.4truecm}
\end{figure*}

Comparisons of Teukolsky-based and EOB waveforms are performed 
once they are aligned in time and phase. Such an alignment guarantees that the fitting factor
is maximized over time and phase of coalescence in a matched filtering calculation with white 
noise~\cite{Buonanno:2009qa}. The alignment procedure depends rather sensitively on the alignment window chosen. 
We choose to align the waveforms in the low-frequency regime, i.e., in the interval $t \in [0,64] \lambda_{\GW}$, where $\lambda_{\GW}$ is the GW wavelength, 
$t \simeq (0,0.006 M)$ [$t \simeq (0,0.013 M)$] months for system-I [system-II], to a level of $10^{-10}$ [$10^{-6}$] rads in the phase for system-I [system-II]. 
We have checked that choosing any interval window of width $< 2^{9} \lambda_{\GW}$ changes the final dephasing by less than $10^{-3}$ rads and the relative
amplitude difference by less than $10^{-6}$. 

In the left panel of Fig.~\ref{fig:plot2} we plot the absolute value of the phase difference, or {\it{dephasing}}, between the dominant
$h_{22}$ mode of the Teukolsky-based and EOB waveforms as a function of time in units of months.  We find that after 2-years the dephasing is $\sim
40$ (3000) rads for system-I (system-II) when using the EOB-model with Taylor-flux (not shown in the figure)~\cite{Boyle:2008ge}, 
a result in qualitative agreement with previous investigations~\cite{Mandel:2008bc}. 
The EOB model with uncalibrated Pad\'e-flux at 5.5PN has a dephasing of 
$\sim 5$ ($530$) rads for system-I (system-II), 
which can be reduced to $\sim 0.1$ ($0.01$) rads if we employ the calibrated Pad\'e-flux at 6.5PN. 
The EOB model with uncalibrated $\rho$-flux at 5.5PN has a dephasing of $\sim 10$ ($530$) 
rads for system-I (system-II),  which can be reduced to $\sim 2$ ($0.8$) rads if we consider the calibrated $\rho$-flux at 6PN.

In the right panel of Fig.~\ref{fig:plot2}, we compare the amplitude of the dominant mode $A_{22}=|h_{22}|$, 
computed in the EOB and Teukolsky frameworks. After 2-years of evolution, both the calibrated Pad\'e- and $\rho$-flux 
EOB models have a disagreement of $\sim 10^{-5}$ for system-I and 
$\sim 2 \times 10^{-3}$ for system-II. 
Such a phase and amplitude agreement is fantastic when one takes into account the 2-year length of observation, 
during which the binary of system-I (system-II) evolves over $\sim 2 \times 10^{6}$ ($\sim 9 \times 10^{5}$) rads.  
Quite interestingly, we find that if we switch on the relative $\nu$ terms in the 3PN EOB Hamiltonian 
Eq.~(\ref{eq:genexp}) (conservative self-force) and in the flux 
(dissipative self-force\footnote{Sometimes {\emph{all}} of the energy loss due to radiation, 
is considered part of the dissipative force (even the $\nu =0$ part), but here we refer 
only to the $\nu$-dependent terms in the flux.}) the dephasing, for the EOB-model with $\rho$-flux at 6PN, increases 
to $\sim 27$ ($6$) rads for system-I (system-II), while the Newtonian normalized amplitude difference 
increases to $4 \times 10^{-4}$ ($2.5 \times 10^{-3}$) for system-I (system-II). We notice that the 
main effect comes from the dissipative self-force, a result consistent with~\cite{Pound:2007th} for circular
orbits (see e.g.~\cite{Blanchet:2009sd,barack_sago,Damour:2009sm} for more details on the PN self-force).

We also compare the strongest higher harmonics using the EOB model with Pad\'e-flux at 6.5PN. 
In the case of the $(\ell,m)=(3,3)$ and $(\ell,m)=(4,4)$ modes we find dephasings of $\sim 0.14$ ($0.07$) and $\sim 0.18$ ($0.09$) 
rads, and normalized amplitude differences of $\sim 6 \times 10^{-5}$ ($4 \times 10^{-3}$) and 
$\sim 3 \times 10^{-4}$ ($9 \times 10^{-3}$), for system-I (system-II). These dephasings are comparable to
those found for the $(\ell,m)=(2,2)$ mode because in both frameworks the GW phase (and frequency) can be computed 
directly from the orbital phase (and frequency), up to errors of less than $\sim 1$ rad over a 2-year 
integration. As a consequence, the above comparisons are almost entirely governed by the trajectories of the 
test particle. Finally, we find that higher harmonics contribute significantly less to the signal-to-noise ratio 
relative to the $(2,2)$ mode. In particular, we computed the signal-to-noise ratio 
averaged over beam-pattern functions with a noise spectral density that includes white-dwarf confusion noise. 
Including up to $\ell = 5$ ($\ell = 7$) for system-I (system-II) guarantees a recovery of $97 \%$ of the total signal-to-noise ratio, 
with the $\ell = m$ modes the most dominant. 

{\it{Data Analysis Implications and Discussion}}. The above results have demonstrated that the EOB framework can be used to model 
EMRIs for LISA data analysis purposes, with the advantage of allowing for the consistent inclusion of both dissipative and
conservative PN self-force terms. In addition, such terms allow the construction of waveforms from intermediate-mass ratio inspirals, 
where first-order BH perturbation theory is expected to fail. The comparisons made here, however, serve only as a proof-of-principle,
as one must now generalize the formalism to more generic spinning EMRIs, and more complicated orbital geometries.

The EOB framework also allows us to provide, for the first time, a metric-based estimate of the number of templates
needed for EMRI systems in LISA data analysis~\cite{Owen:1995tm,Damour:1997ub}. 
As a coherent 2-year integration in the search of EMRIs is computationally prohibitive, 
a hierarchical search that collects power from coherent searches of shorter segments was proposed
in~\cite{2004CQGra..21S1595G}. The maximum segment length set by computational 
limits in such a hierarchical search is estimated to be less than $2$ months. For a 2-month evolution, 
we estimate that one requires less than $10^{7}$ EOB templates to 
cover the template bank with a minimal match of $0.97$ in the total
mass range $(10^{5}$--$10^{6}) M_{\odot}$ and mass ratio range $(10^{-4}$--$10^{-5})$.

{\it{Acknowledgments}}.
We are grateful to F.~Pretorius and E.~Poisson for comments and to W.~Throwe for computational assistance.
NY, AB and YP, and SAH acknowledge support from the NSF grants PHY-0745779, 
PHY-0603762, PHY-0903631, and PHY-0449884; AB, SAH and MCM also acknowledge 
support from NASA grants NNX09AI81G, NNX08AL42G and NNX08AH29G.




\end{document}